\documentclass{article}
\usepackage{spconf,amsmath,graphicx}
\usepackage{booktabs}
\usepackage{multirow}
\usepackage{xcolor}
\usepackage{amssymb}

\title{End-to-End speech recognition contextualization with Large Language Models}
\name{Egor Lakomkin, Chunyang Wu, Yassir Fathullah$^{\dagger}$\thanks{Work done during internship at Meta AI.}, Ozlem Kalinli, Michael L. Seltzer, Christian Fuegen}
\address{Meta AI}

\begin{document}
%
\maketitle
\begin{abstract}

In recent years, Large Language Models (LLMs) have garnered significant attention from the research community due to their exceptional performance and generalization capabilities. In this paper, we introduce a novel method for contextualizing speech recognition models incorporating LLMs. 
Our approach casts speech recognition as a mixed-modal language modeling task based on a pretrained LLM. We provide audio features, along with optional text tokens for context, to train the system to complete transcriptions in a decoder-only fashion. As a result, the system is implicitly incentivized to learn how to leverage unstructured contextual information during training. 
Our empirical results demonstrate a significant improvement in performance, with a 6\% WER reduction when additional textual context is provided. Moreover, we find that our method performs competitively and improve by 7.5\% WER overall and 17\% WER on rare words against a baseline contextualized RNN-T system that has been trained on more than twenty five times larger speech dataset. Overall, we demonstrate that by only adding a handful number of trainable parameters via adapters, we can unlock contextualized speech recognition capability for the pretrained LLM while keeping the same text-only input functionality. 
\end{abstract}
\begin{keywords}
contextual biasing, large language models, speech recognition
\end{keywords}
\vspace{-3mm}
\section{Introduction}
\label{sec:intro}

\par In recent years, there has been growing interest in Large Language Models (LLMs) due to their remarkable efficacy in enhancing performance in tasks like question answering and summarization, surpassing specialized models \cite{touvron2023llama, openai2023gpt4}. LLMs are trained on vast quantities of text data, thereby encapsulating a wealth of world knowledge within the network. This accumulated knowledge and contextual understanding prove to be particularly beneficial in the field of Automatic Speech Recognition (ASR), especially when additional context surrounding an utterance is available beyond the audio alone. For example, video titles and descriptions can provide insights into the topic of the video or offer clues about named entities that might be mentioned \cite{DBLP:conf/interspeech/JainKMZMS20, DBLP:conf/interspeech/LeJKKSMCSFKSS21}. Such contextual information can assist in disambiguating challenging pronunciations, as certain words, domain-specific terms, or named entities can often be inferred from context alone.
Traditional approaches to ASR contextualization \cite{DBLP:conf/interspeech/LeJKKSMCSFKSS21, DBLP:conf/interspeech/JainKMZMS20, DBLP:conf/icassp/SathyendraMCLSS22, DBLP:conf/slt/PundakSPKZ18} operate at the token or phrase level, employing techniques like biasing with weighted finite state transducers (WFSTs) or using specialized attention networks. These are typically either incorporated during the decoding stage or trained as separate components. Consequently, contextualization significantly improves the ASR system's ability to recognize named entities or specialized in-domain terms. However, there are some limitations to these approaches:

- The biasing is limited towards individual phrases or words, as opposed to contextualizing based on external information as a whole (for example, topic-based biasing).

- The biasing strength is usually controlled via a hyperparameter or requires specialized architectural changes and training procedures to ensure the system is not overbiased.

- Some of the contextualization methods influence only the decoder state without interacting with the encoder directly.

\begin{figure}[t]

  \centering
  \centerline{\includegraphics[width=8.5cm]{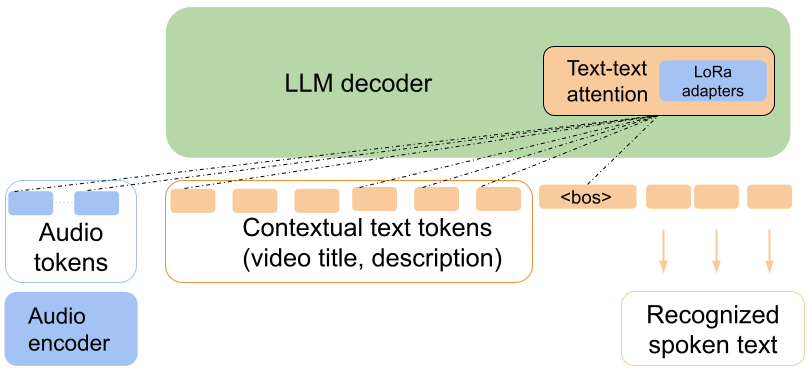}}

\caption{A speech recognition model with mixed-modal context consisting of audio and optional text tokens based on a pretrained LLM backbone. Speech encoder and LLM decoder are both initially pretrained. The LLM weights are frozen (orange blocks), while audio encoder and LoRa adapters are fine-tuned during training (blue blocks).}
\label{fig:decoder_only_architecture}
\vspace{-2mm}
\end{figure}

\par In this work, we propose a Speech LLaMA - a  decoder-only architecture inspired by recent developments in LLMs tailored towards speech recognition. It is trained to use the contextual information end-to-end without any additional hyperparameters. Specifically, 1) we prepend the whole available textual context as a prompt to an ASR system along with audio tokens. 

\noindent The Speech LLaMA hence have the full flexibility to look back and cross-corellate the contextual text tokens and the acoustic tokens when decoding the next spoken word. And 2) we employ the publicly available 7B LLaMA LLM~\cite{touvron2023llama} as a pretrained decoder for the Speech LLaMA. This simplifies the overall design of a contextual ASR as speech recognition can be considered as mixed-modal language model with next-token prediciton. Our intuition behind this is the pretrained LLMs already distill the linguistic information which should be particularly useful when reasoning which part of the context is relevant given the utterance. Our results on a competitive benchmark suggest a feasibility of this modelling approach.

\begin{table*}[t]
\centering
\caption{Evaluation results of Speech LLaMA compared to large-scale RNN-T baseline on English speech data. We report overall WER and Rare WER. Rare WER specifically focuses on the accuracy of recognizing rare words in the dataset.}
\label{table:model_performance}
\vspace{5mm}
\begin{tabular}{llcccccccc} 

\toprule
Model & \multirow{2}{*}{\shortstack{Speech\\ data (h)}} & \multirow{2}{*}{\shortstack{Trainable\\ params (M)}} &  \multicolumn{2}{c}{Context presence} & WER (\%) & SUB & INS & DEL & Rare WER (\%)\\
\cmidrule(lr){4-5} 
      & &  & Training & Evaluation \\
\midrule
1B RNN-T \cite{Xiao2021ScalingAI} & 4M  & 1000 & - & - & 12.34  & 6.53 & 3.21 & 2.60  & 30.80 \\
1B RNN-T \cite{Xiao2021ScalingAI} & 4M  & 1000 & - & $\checkmark$ & 12.13  & 6.23 & 3.05 & 2.85 & 28.96 \\
Speech LLaMa & 150k  & 130 & - & -  & 11.70  & 6.09 & 3.20 & 2.38 & 27.33  \\
Speech LLaMa & 150k  & 130 & $\checkmark$  & - & 11.98  & 6.28 & 3.07 & 2.63 & 28.64 \\
Speech LLaMa & 150k  & 130 & $\checkmark$  & $\checkmark$  & \textbf{11.22} & 5.76 & 3.14 & 2.32 & \textbf{23.88} \\
\bottomrule
\end{tabular}
\vspace{-2mm}
\end{table*}

\vspace{-2mm}
\section{Related Work}
\label{sec:format}

\par There have been several works on speech recognition models contextualization including deep and shallow biasing \cite{DBLP:conf/interspeech/ZhaoSRRBLP19, DBLP:conf/interspeech/LeJKKSMCSFKSS21}. Le et al. \cite{DBLP:conf/interspeech/LeJKKSMCSFKSS21} introduced a weighted finite state transducer (WFST) composed from biasing strings which is attached dynamically during decoding and the scores of the RNN-T system and biasing WFST are interpolated. The advantage of such approaches is that they could be attached to any system after the training is completed. Another line of research is deep biasing methods that incorporate contextualization end-to-end during the model training \cite{9383560, DBLP:conf/interspeech/JainKMZMS20, DBLP:conf/slt/PundakSPKZ18, DBLP:journals/corr/abs-2305-05271, xu2023adaptive, DBLP:conf/icassp/SathyendraMCLSS22}.  A common limitation of these approaches is that the bias on the phrase level, rather than providing on the full context available.  In addition, these approaches require a specialized biasing modules added to the main ASR architecture.

\par In parallel to this reseach several approaches were presented incorporating LLMs for speech related tasks. Wu at al. \cite{wu2023decoderonly} incorporated LLaMA LLM for speech translation by concatenating a textual prompt (\textit{"Translate audio to language X"}) with audio representations. AudioPalm \cite{rubenstein2023audiopalm} model was proposed mixing audio and text tokens for speech-to-text and speech to speech tasks. Fathullah et al. \cite{fathullah2023prompting} presented results on enabling speech recognition capabilities for LLaMA model on the multi-lingual data. Recently a Whisper model \cite{pmlr-v202-radford23a} incorporated a biasing approach, where the previous segment's transcription was added to the prompt for the long-form speech recognition. In difference to their work, we bias the system on the unstructed and sometimes unrelated textual context as not always video title and description match the context of speech. 

\vspace{-3mm}
\section{Experimental setup}
\label{sec:pagestyle}

\textbf{Model}: 
Figure~\ref{fig:decoder_only_architecture} illustrates the overview of our proposed model.
This speech LLM architecture consists of two main blocks: audio encoder and text decoder. The audio encoder firstly applies four downsampling blocks resuling in 16x time reduction of audio representations. After that a stack of Conformer \cite{DBLP:conf/interspeech/GulatiQCPZYHWZW20} blocks with rotary positional embeddings \cite{DBLP:journals/corr/abs-2104-09864} are applied with hidden dimensionality of 512 and kernel size of 9.
At the end we add an additional downsampling block. As a result the decoder observes audio tokens sampled every 320ms with dimensionality of size 4,096. We pretrained the audio encoder with Connectionist Temporal Classification \cite{DBLP:conf/icml/GravesFGS06} criterion for 300k training steps on the same training data. We used a pretrained 7B LLaMA (v1) \cite{touvron2023llama} as a decoder. To adapt text-only LLaMA to speech recognition task, we have added Low-Rank Adapters \cite{DBLP:conf/iclr/HuSWALWWC22} to query, key, value, and output projection matrices in the self-attention layer of every decoder layer while keeping the rest of LLM parameters frozen throughout the training. We used the following LoRa parameters: rank of size 32, dropout rate of 5\%, and 0.05 scaling parameter. Overall LoRa parameters add 30 million trainable parameters to the LLM decoder and the rest 6.7 billion are kept frozen.

We used 80 dimensional log Mel features computed every 10ms with a window
of 25ms. SpecAugment \cite{Park_2019} with two frequency masks of width 27 and ten time masks with maximum width of 4\% of the length of an utterance.  We trained our models for 200,000 updates with mixed precision, linearly increasing the learning rate to $5\text{e-}4$ in the first 20,000
updates and exponentially decaying to $1\text{e-}5$
over the remaining updates. We use Adam with parameters $\beta 1$ = 0.9, $\beta 2$ = 0.98, weight decay = $1\text{e-}5$
and clip the gradient norm to 1. Our model is trained with 128 A100 GPUs for 3 days using Fairseq library \cite{ott-etal-2019-fairseq}.

\textbf{Data}: The models are trained on an in-house dataset that was de-identified with no personally identifiable information (PII) derived from public Facebook and Instagram videos. The data was further augmented with two distortion methods: speed perturbation \cite{DBLP:conf/interspeech/KoPPK15} and randomly sampled additive background noise. 
For evaluation, we have sampled 3,200 videos comprising around 34 hours of speech that have context of at least 100 characters length with at least one non-frequent word from the context occurs in the transcription.

\textbf{Metrics}: To evaluate our models, we report both the overall Word Error Rate (WER) and \textit{Rare WER}, which considers only rare words. A word is considered rare if it does not occur in the 90\% percentile of the most frequent words estimated on training data.

\textbf{Textual context}: Similar to Xiao et al. \cite{Xiao2021ScalingAI} we incorporate video title and video description as an external context. We perform basic text post-processing like unicode character normalization and removal of all non-ascii symbols. Overall approximately 25\% of videos from supervised video dataset have non-empty text context.
When video title or description are present, we first concatenate and then tokenize them with the LLaMA tokenizer. After that, we prepend the \textit{\textless bos\textgreater} token with the textual tokens. When both video title and descriptions are missing, the input corresponds to a traditional ASR setup without contextual information.  The cross-entropy loss is masked for the contextual tokens and only computed for spoken tokens.
In these experiments we limit the textual content to a maximum of 50 tokens for computational reasons. If the context is longer than the threshold, we perform a random crop of size 50 during training and crop the leading tokens during inference.

\textbf{Baseline}: As a baseline we used a transformer based RNN-T system with one billion parameters  \cite{Xiao2021ScalingAI}, which is trained on four million hours of supervised and semi-supervised speech data. The RNN-T system architecture consists of 60 transformer layers in the encoder and 3 LSTM layers in the decoder. For contextualization it uses an WFST biasing method with neural language modelling shallow fusion \cite{DBLP:conf/interspeech/LeJKKSMCSFKSS21}, where the biasing FST is composed from video title and description. We are using exactly the same contextual information during decoding for the RNN-T baseline and our Speech LLaMA.

\vspace{-1em}
\section{Results}
\label{sec:typestyle}

Table \ref{table:model_performance} presents a summary of our decoding results on the evaluation set. We compare the Speech LLaMA against the offline RNN-T 1b model, considering two scenarios: with and without presenting contextualization information during decoding. The WER scores obtained for these scenarios using RNN-T are 12.34\% and 12.13\% respectively. Contextual biasing resuts in a relative WER reduction of approximately 1.7\%. 

Even without the use of contextual information during training and evaluation, Speech LLaMA achieves a WER of 11.70\%, a relative reduction of 5.2\% over the RNN-T system trained on much more data.

By incorporating context during training and evaluation, we achieve a significant improvement reaching an overall WER of 11.22\% and resulting in 17\% relative improvement in Rare WER, surpassing the performance of the RNN-T model with contextualization.

It is worth noting that when we evaluate the Speech LLaMA  trained with context but do not provide the context during inference, we obtain a WER of 11.98\%. This corresponds to a slight WER gap compared to the model trained without context. We leave to address this minor performance difference to the future work, where adding a certain jitter to the context may improve the generalization of a model towards presence of the context.

\subsection{Ablation studies}

\subsubsection{Context sensitivity}
\par To better understand how the model learns to use the context, we studied how receptive the model is to context perturbations. For this we tried a few ways to modify the prompt and measure its effect on the decoding. Specifically, we experimented with:
\begin{enumerate}
  \item Replacing the actual context with words randomly sampled from the training data.
  \item Replacing the context with the ground truth words.  We filter out frequent words in this experiment as we assume that the model should not have significant issues in transcribing them. We expect a significant reduction of WER if the model is capable of copy-pasting the words from the context.
  \item Replacing the contextual words with phonetical respellings of the words that appear in the transcripts. Our intuition is that such replacements are particularly challenging for the model and we should expect a bigger WER change compared to random substitutions. To generate re-spellings we employed a G2G \cite{9054257} model. For every rare word in the ground truth we sample an alternative spelling from the G2G model and add it to the context. For example, if the word \textit{ball} is present in the context and ground truth we replace it by \textit{bawl} and use that as context instead of the original token.
  \item In addition to the previous perturbation we probe appending a similar sounding word to the context (e.g. both tokens \textit{ball} and \textit{bawl} will be present in the context). This tests the ability of an ASR system to disambiguate the actual spoken word given a competitive word in context.
\end{enumerate}

\vspace{-4mm}
\begin{table}[ht]
\centering
\caption{WER under different context perturbations during decoding stage.}
\label{table:wer_noise}
\vspace{5mm}
\begin{tabular}{lcc} 
\toprule
Context noise         & WER (\%) & Rare WER (\%) \\
\midrule
(Original context)          & 11.22   & 23.88    \\
(Remove all context) & 11.98  & 28.64 \\
Random       & 12.07   & 28.85  \\ 
Respellings  & 11.89   &  28.31   \\  
Respellings (append)  & 11.46  & 25.59     \\  
Ground Truth  & 10.50   & 19.54   \\  

\bottomrule
\end{tabular}
\end{table}

We present our results in Table \ref{table:wer_noise}. We note that replacing the whole context with random words sampled from the training data results in only a marginal difference in WER compared to removing all external context (11.98\% vs. 12.07\%). This indicates that the model is robust against some contextual noise and can distinguish relevant from irrelevant context. Substituting rare words that match both the context and the ground truth with G2G respellings results in a significant drop in WER (11.22\% $\rightarrow$ 11.89\%), almost matching with not using any context. This hints that the majority of gains observed are due to the model being able to copy certain words from the context. In contrast, when we instead of replacing the matching contextual word rather append a competing similar-sounding word, we observe a smaller WER drop (11.22\% $\rightarrow$ 11.46\%). This indicates that the model does not necessarily get confused by similarly pronounced words with different meanings. Furthermore, when we take the rare words from the ground truth into the context, the WER improves to 10.50\% (6\% relative change) and Rare WER improves by 18\% relative. This further proves the ability of the model to utilize contextual information when present in order to better recognize rare entities.

\vspace{-3mm}
\begin{table}[h]
\centering
\caption{Impact of the context masking structure on the WER.}
\label{table:masking_performance}
\vspace{3mm}
\begin{tabular}{lc} 
\toprule
Masking & WER (\%) \\
\midrule
Causal    & 11.22     \\
Full-Mask  & 11.15    \\ 
\bottomrule
\end{tabular}
\vspace{-6mm}
\end{table}

\subsubsection{Causal vs Full Masking}

\par Traditionally causal masking is used in all self-attention layers for decoder-only language models to prevent future information leakage. However for offline speech recognition we have full audio and text context observed at the time of decoding and only transcription tokens are necessary to be masked causally. In this section we experiment the impact of applying causal masking on all input tokens and contrast it with applying full mask on the text and audio context followed by causal masking on transcription tokens. While the audio representations are fully contextualized already, we hypothesize that textual tokens may benefit from full masking.
\par We present our results in Table \ref{table:masking_performance}. The full-mask shows only marginally better WER then causal masking (improving from 11.22\% $\rightarrow$ to 11.15\%). This comes at a cost as efficient self-attention implementations are currently tailored towards causal masking (Flash-Attention v2) and using a custom masking slows down training by 10\%. 

\subsubsection{Decoder-only vs Cross-attention}

Furthermore, we compared the decoder-only approach to a traditional encoder-decoder model by converting the Speech LLM architecture to Listen-Attend-Spell architecture \cite{DBLP:journals/corr/ChanJLV15}. To achieve that, instead of concatenating audio and text tokens we treated them separaterely. We added trainable cross-attention matrices to every LLM decoder layer.
Table 3 presents the results of this study. We observed that the two approaches perform similarly, with only minor improvement for the Encoder-Decoder architecture (11.22\% $\rightarrow$ 11.18\%). This indicates that the decoder-only approach is a viable and straightforward method for performing ASR with or without contextualization.

However, one limitation of the decoder-only approach is the quadratic attention complexity, which can impose restrictions on the overall sequence length. This limitation becomes significant as the context grows. To address this issue, we can employ techniques such as lower precision training (8 or 4 bits) and linear attention approximation methods \cite{DBLP:conf/icml/DettmersZ23, ding2023longnet}.

\begin{table}[t]
\centering
\caption{Performance comparison of decoder-only Speech LLM and cross-attention Speech LLM. }
\label{table:decoder_performance}
\vspace{3mm}
\begin{tabular}{lc} 
\toprule
Decoder  & WER (\%)  \\
\midrule
 Decoder-only  & 11.22  \\
Encoder-decoder & 11.18  \\
\bottomrule
\end{tabular}
\end{table}

\vspace{-2mm}
\section{Conclusions and Future Work}
\label{sec:typestyle}

In this work, we have presented to our knowledge the first results on utilizing pretrained LLMs to leverage contextual information in order to improve speech recognition. We have demonstrated that with a simple decoder-only architecture we can condition the ASR output on the unstructured text.  Our approach shows superior performance against a strong baseline, proving the feasability of the proposed method at scale. End-to-end contextualization via text promping with LLMs shows better context utilization compared to our strong RNN-T based baselines. In addition, our ablation studies show that the system is robust to noise perturbations and shows abilities to perform a phonetic disambiguation. As part of the future work, we plan to extend the methods towards long context and other modalities.

\bibliographystyle{IEEEbib}
\bibliography{refs}

\begin{thebibliography}{10}

\bibitem{touvron2023llama}
Hugo Touvron, Thibaut Lavril, Gautier Izacard, et~al.,
\newblock ``Llama: Open and efficient foundation language models,'' 2023.

\bibitem{openai2023gpt4}
OpenAI,
\newblock ``Gpt-4 technical report,'' 2023.

\bibitem{DBLP:conf/interspeech/JainKMZMS20}
Mahaveer Jain, Gil Keren, Jay Mahadeokar, Geoffrey Zweig, Florian Metze, and
  Yatharth Saraf,
\newblock ``Contextual {RNN-T} for open domain {ASR},''
\newblock in {\em INTERSPEECH}. 2020, pp. 11--15, {ISCA}.

\bibitem{DBLP:conf/interspeech/LeJKKSMCSFKSS21}
Duc Le, Mahaveer Jain, Gil Keren, Suyoun Kim, et~al.,
\newblock ``Contextualized streaming end-to-end speech recognition with
  trie-based deep biasing and shallow fusion,''
\newblock in {\em INTERSPEECH}, 2021, pp. 1772--1776.

\bibitem{DBLP:conf/icassp/SathyendraMCLSS22}
Kanthashree~Mysore Sathyendra, Thejaswi Muniyappa, Feng{-}Ju Chang, et~al.,
\newblock ``Contextual adapters for personalized speech recognition in neural
  transducers,''
\newblock in {\em ICASSP}. 2022, pp. 8537--8541, {IEEE}.

\bibitem{DBLP:conf/slt/PundakSPKZ18}
Golan Pundak, Tara~N. Sainath, et~al.,
\newblock ``Deep context: End-to-end contextual speech recognition,''
\newblock in {\em 2018 {IEEE} Spoken Language Technology Workshop}. 2018, pp.
  418--425, {IEEE}.

\bibitem{Xiao2021ScalingAI}
Alex Xiao, Weiyi Zheng, Gil Keren, et~al.,
\newblock ``Scaling asr improves zero and few shot learning,''
\newblock in {\em INTERSPEECH}, 2021.

\bibitem{DBLP:conf/interspeech/ZhaoSRRBLP19}
Ding Zhao, Tara~N. Sainath, David Rybach, Pat Rondon, et~al.,
\newblock ``Shallow-fusion end-to-end contextual biasing,''
\newblock in {\em INTERSPEECH}. 2019, pp. 1418--1422, {ISCA}.

\bibitem{9383560}
Duc Le, Gil Keren, Julian Chan, et~al.,
\newblock ``Deep shallow fusion for rnn-t personalization,''
\newblock in {\em 2021 IEEE Spoken Language Technology Workshop (SLT)}, 2021,
  pp. 251--257.

\bibitem{DBLP:journals/corr/abs-2305-05271}
Xuandi Fu, Kanthashree~Mysore Sathyendra, Ankur Gandhe, et~al.,
\newblock ``Robust acoustic and semantic contextual biasing in neural
  transducers for speech recognition,''
\newblock {\em CoRR}, vol. abs/2305.05271, 2023.

\bibitem{xu2023adaptive}
Tianyi Xu, Zhanheng Yang, Kaixun Huang, et~al.,
\newblock ``Adaptive contextual biasing for transducer based streaming speech
  recognition,'' 2023.

\bibitem{wu2023decoderonly}
Jian Wu, Yashesh Gaur, et~al.,
\newblock ``On decoder-only architecture for speech-to-text and large language
  model integration,'' 2023.

\bibitem{rubenstein2023audiopalm}
Paul~K. Rubenstein et~al.,
\newblock ``Audiopalm: A large language model that can speak and listen,''
\newblock 2023.

\bibitem{fathullah2023prompting}
Yassir Fathullah, Chunyang Wu, Egor Lakomkin, et~al.,
\newblock ``Prompting large language models with speech recognition
  abilities,'' 2023.

\bibitem{pmlr-v202-radford23a}
Alec Radford, Jong~Wook Kim, et~al.,
\newblock ``Robust speech recognition via large-scale weak supervision,''
\newblock in {\em ICML}. 23--29 Jul 2023, vol. 202 of {\em Proceedings of
  Machine Learning Research}, pp. 28492--28518, PMLR.

\bibitem{DBLP:conf/interspeech/GulatiQCPZYHWZW20}
Anmol Gulati, James Qin, et~al.,
\newblock ``Conformer: Convolution-augmented transformer for speech
  recognition,''
\newblock in {\em INTERSPEECH}. 2020, pp. 5036--5040, {ISCA}.

\bibitem{DBLP:journals/corr/abs-2104-09864}
Jianlin Su, Yu~Lu, et~al.,
\newblock ``Roformer: Enhanced transformer with rotary position embedding,''
\newblock {\em CoRR}, vol. abs/2104.09864, 2021.

\bibitem{DBLP:conf/icml/GravesFGS06}
Alex Graves et~al.,
\newblock ``Connectionist temporal classification: labelling unsegmented
  sequence data with recurrent neural networks,''
\newblock in {\em ICML}. 2006, vol. 148 of {\em {ACM} International Conference
  Proceeding Series}, pp. 369--376, {ACM}.

\bibitem{DBLP:conf/iclr/HuSWALWWC22}
Edward~J. Hu, Yelong Shen, et~al.,
\newblock ``Lora: Low-rank adaptation of large language models,''
\newblock in {\em ICLR}. 2022, OpenReview.net.

\bibitem{Park_2019}
Daniel~S. Park, William Chan, Yu~Zhang, Chung-Cheng Chiu, Barret Zoph, Ekin~D.
  Cubuk, and Quoc~V. Le,
\newblock ``Specaugment: A simple data augmentation method for automatic speech
  recognition,''
\newblock {\em INTERSPEECH}, Sep 2019.

\bibitem{ott-etal-2019-fairseq}
Myle Ott, Sergey Edunov, Alexei Baevski, et~al.,
\newblock ``fairseq: A fast, extensible toolkit for sequence modeling,''
\newblock in {\em ACL (Demonstrations)}, Minneapolis, Minnesota, June 2019, pp.
  48--53, Association for Computational Linguistics.

\bibitem{DBLP:conf/interspeech/KoPPK15}
Tom Ko, Vijayaditya Peddinti, et~al.,
\newblock ``Audio augmentation for speech recognition,''
\newblock in {\em {INTERSPEECH}}. 2015, pp. 3586--3589, {ISCA}.

\bibitem{9054257}
Duc Le, Thilo Koehler, Christian Fuegen, and Michael~L. Seltzer,
\newblock ``G2g: Tts-driven pronunciation learning for graphemic hybrid asr,''
\newblock in {\em ICASSP}, 2020, pp. 6869--6873.

\bibitem{DBLP:journals/corr/ChanJLV15}
William Chan, Navdeep Jaitly, Quoc~V. Le, and Oriol Vinyals,
\newblock ``Listen, attend and spell,''
\newblock {\em CoRR}, vol. abs/1508.01211, 2015.

\bibitem{DBLP:conf/icml/DettmersZ23}
Tim Dettmers and Luke Zettlemoyer,
\newblock ``The case for 4-bit precision: k-bit inference scaling laws,''
\newblock in {\em ICML}. 2023, vol. 202, pp. 7750--7774, {PMLR}.

\bibitem{ding2023longnet}
Jiayu Ding, Shuming Ma, et~al.,
\newblock ``Longnet: Scaling transformers to 1,000,000,000 tokens,'' 2023.

\end{thebibliography}

\end{document}